\documentclass[twocolumn,floatfix,prb,eqsecnum,superscriptaddress]{revtex4}
\usepackage{graphicx}
\usepackage{amsmath}
\usepackage{amssymb}
\usepackage{bm}

\usepackage{color}

\begin{document}

\title{Phase-locked magnetoconductance oscillations as a probe of Majorana edge states}
\author{M. Diez}
\affiliation{Instituut-Lorentz, Universiteit Leiden, P.O. Box 9506, 2300 RA Leiden, The Netherlands}
\author{I. C. Fulga}
\affiliation{Instituut-Lorentz, Universiteit Leiden, P.O. Box 9506, 2300 RA Leiden, The Netherlands}
\author{D. I. Pikulin}
\affiliation{Instituut-Lorentz, Universiteit Leiden, P.O. Box 9506, 2300 RA Leiden, The Netherlands}
\author{M. Wimmer}
\affiliation{Instituut-Lorentz, Universiteit Leiden, P.O. Box 9506, 2300 RA Leiden, The Netherlands}
\author{A. R. Akhmerov}
\affiliation{Department of Physics, Harvard University, Cambridge, Massachusetts 02138 USA}
\author{C. W. J. Beenakker}
\affiliation{Instituut-Lorentz, Universiteit Leiden, P.O. Box 9506, 2300 RA Leiden, The Netherlands}

\date{January 2013}
\begin{abstract}
We calculate the Andreev conductance of a superconducting ring interrupted by a flux-biased Josephson junction, searching for electrical signatures of circulating edge states. Two-dimensional pair potentials of spin-singlet \textit{d}-wave and spin-triplet \textit{p}-wave symmetry support, respectively, (chiral) Dirac modes and (chiral or helical) Majorana modes. These produce $h/e$-periodic magnetoconductance oscillations of amplitude $\simeq (e^{2}/h)N^{-1/2}$, measured via an $N$-mode point contact at the inner or outer perimeter of the grounded ring. For Dirac modes the oscillations in the two contacts are independent, while for an unpaired Majorana mode they are phase locked by a topological phase transition at the Josephson junction.
\end{abstract}
\maketitle

\section{Introduction}
\label{intro}

Two-dimensional superconductors can support propagating edge states that are not localized by disorder for topological reasons,\cite{Vol97,Sen99,Rea00} as a superconducting analogue of the metallic edge states in the quantum Hall effect or quantum spin-Hall effect.\cite{Has10,Qi11} Unlike the dispersionless ``flat band'' edge states of nodal superconductors,\cite{Kas00} which leave a strong signature in the density of states, the propagating edge states have not yet been observed. They have been predicted in a variety of materials --- including strontium ruthenate,\cite{Mac03} heavily doped graphene,\cite{Nan12} and topological insulators on a superconducting substrate.\cite{Fu08}

The symmetry-based classification of topological superconductors lists three types of propagating edge states:\cite{Ryu10} chiral Dirac modes, and chiral or helical Majorana modes (see Table \ref{tab:table1}). A spin-singlet superconductor with $d_{x^2-y^2}+id_{xy}$ orbital pairing supports edge states that propagate in one direction only (chiral) and are not selfconjugate (Dirac fermions). For spin-triplet $p_x+ip_{y}$ pairing the edge states are chiral and selfconjugate (Majorana fermions). Counterpropagating (helical) Majorana modes are also stable against localization,\cite{Qi09,Tan09,Sat09} unlike counterpropagating Dirac modes.

\begin{table}[tb]
\centering
\begin{tabular}{  c | c | c }
pair potential & edge state & symmetry class\\ \hline
singlet, $d_{x^2-y^2}+id_{xy}$ & chiral Dirac & C \\
triplet, $p_x+ip_{y}$ & chiral Majorana & D \\
triplet, $p_x\pm ip_{y}$ & helical Majorana & DIII
\end{tabular}
\caption{The three types of propagating edge states in a two-dimensional topological superconductor.}
\label{tab:table1}
\end{table}

The topological protection allows for correlations in the electrical current measured at distant points on the boundary connected by an edge state.\cite{Law09,Ser10,Chu11,Liu11,Str11,Li12,Ber12} For example, in an early study of this type, Law, Lee, and Ng considered a superconducting disc deposited on the surface of a three-dimensional topological insulator.\cite{Law09} A chiral Majorana mode is confined to the perimeter of the disc, when the surface outside the superconductor is gapped by a ferromagnet.\cite{Fu08} Two point contacts attached to the perimeter measure a correlated current, mediated by the circulating edge state, and dependent on the number of magnetic vortices inside the disc. It is essential that the contacts share a boundary. If the disc would be replaced by a ring, with one contact at the inner and one at the outer perimeter, then there would be no correlations.

\begin{figure}[tb]
\centerline{\includegraphics[width=1\linewidth]{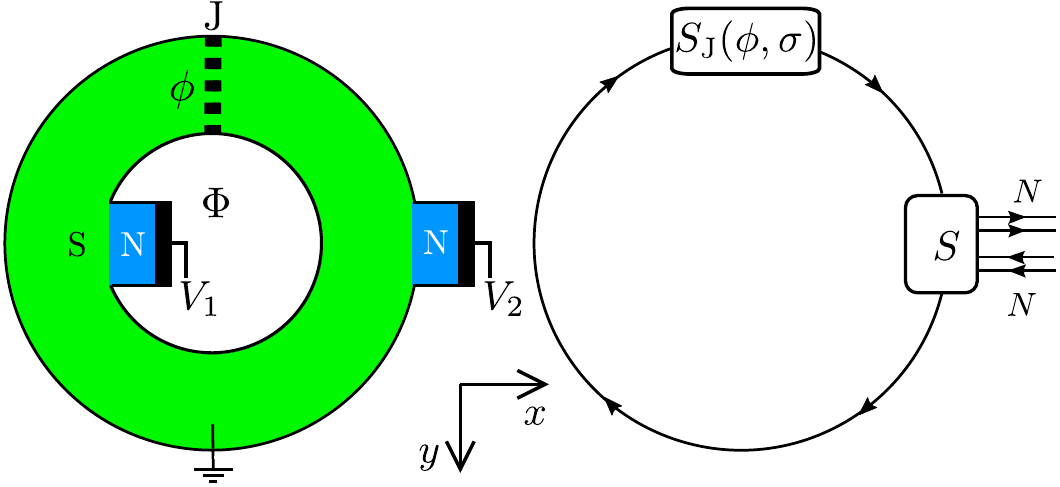}}
\caption{\textit{Left panel:} Superconducting ring (S) containing a weak link (J), forming a flux-biased Josephson junction. A current can be injected into the grounded superconductor from a voltage-biased normal-metal (N) contact at the inner or outer perimeter. The Andreev conductance $G_{n}=I/V_{n}$ is the ratio of the current-to-ground $I$ and the applied voltage $V_{n}$ to contact $n=1,2$ (with $V=0$ for the other contact). \textit{Right panel:} Scattering processes at the outer perimeter, involving the coupling of a chiral edge state to $2N$ incoming and $2N$ outgoing electron-hole modes at the normal metal. This coupling introduces a dependence of the conductance on the phase difference $\phi=(2e/\hbar)\Phi$ across the Josephson junction and on its topological quantum number $\sigma$.}
\label{fig:setup}
\end{figure}

Here we revisit this problem of edge-state mediated correlations in a superconducting ring, to show that the physics changes qualitatively if the ring contains a weak link forming a Josephson junction (see Fig.\ \ref{fig:setup}). The weak link is a one-dimensional subsystem of the two-dimensional topological superconductor, with its own topological phase transition.\cite{Teo10} Since magnetic flux $\Phi$ can enter into the ring along the junction, there is no flux quantization and we can ask for the $\Phi$-dependence of the conductances $G_{1}$ and $G_{2}$ measured between the grounded superconductor and either the inner or the outer perimeter. Dirac and Majorana modes both produce $h/e$-periodic oscillations in the conductances, but only Majorana modes can correlate the oscillations in $G_{1}$ and $G_{2}$. The mechanism by which the inner and outer perimeter communicate is a closing of the excitation gap when the Josephson junction undergoes a one-dimensional topological phase transition. The same conclusion was reached recently by Wieder, Zhang, and Kane.\cite{Wie13}

The outline of the paper is as follows. In the next section we formulate the scattering problem of Andreev reflection from the perimeter of a superconducting ring, to obtain a general formula for the electrical conductance. The conductance $G(\Phi,\sigma)$ depends in general both on the enclosed flux $\Phi$ and on the $\mathbb{Z}_{2}$ topological quantum number $\sigma$ of the Josephson junction. A topological phase transition switches $\sigma$ between $+1$ and $-1$, resulting in a jump $\delta G$ of the conductance. In Sec.\ \ref{chiraledge} we calculate the probability distribution $P(\delta G)$ in an ensemble of disordered rings, using the method of random-matrix theory. These are model-independent results, which take as input only the symmetry class of the topological superconductor. We then turn, in Sec.\ \ref{sec:models}, to specific model Hamiltonians in each symmetry class. The numerical results for these models are discussed in Sec.\ \ref{discuss} to arrive at a set of experimentally observable signatures of (1) the presence of circulating edge states and (2) their Majorana or Dirac fermionic nature.

\section{Scattering formula for the conductance}
\label{scatteringG}

We formulate the scattering problem for a superconducting ring interrupted by a Josephson junction, enclosing a magnetic flux $\Phi$. The ring is contacted at the inner or outer perimeter by a normal-metal electrode. Far away from any gap closings there is no transmission between the inner and outer perimeter, so we can treat these two scattering problems independently.  We calculate the Andreev conductance $G$ between the normal metal (N) and the (grounded) superconductor (S). An $h/e$-periodic flux dependence of the conductance serves as a signature of edge states circulating along the ring.

As illustrated in Fig.\ \ref{fig:setup}, the NS interface is described by a scattering matrix $S$, with submatrices $r_{\rm N}$ (reflection back into the normal metal), $t_{\rm edge,N}$, $t_{\rm N,edge}$ (transmission from the normal metal into an edge state, and vice versa), and $t_{\rm edge}$ (transmission along an edge state without entering the normal metal):
\begin{equation}
S=\begin{pmatrix}
r_{\rm N}&t_{\rm N,edge}\\
t_{\rm edge,N}&t_{\rm edge}
\end{pmatrix}.\label{Sdef}
\end{equation}
Incoming and outgoing wave amplitudes at the NS interface are related by
\begin{align}
&a^{\rm out}_{\rm N}=r_{\rm N}a^{\rm in}_{\rm N}+t_{\rm N,edge}a^{\rm in}_{\rm edge}\\
&a^{\rm out}_{\rm edge}=t_{\rm edge,N}a^{\rm in}_{\rm N}+t_{\rm edge}a^{\rm in}_{\rm edge}.\label{aoutaindef}
\end{align}

The scattering matrix $S_{\rm J}$ of the Josephson junction describes how the edge states return back to the NS interface after encircling the ring,
\begin{equation}
a^{\rm in}_{\rm edge}=S_{\rm J}a^{\rm out}_{\rm edge}.\label{SJdef}
\end{equation}
Elimination of the edge state amplitudes gives the relation $a^{\rm out}_{\rm N}=Ra^{\rm in}_{\rm N}$, with the effective reflection matrix of the NS interface,
\begin{equation}
R=r_{\rm N}+t_{\rm N,edge}(1-S_{\rm J}t_{\rm edge})^{-1}S_{\rm J}t_{\rm edge,N}.\label{Rdef}
\end{equation}

The matrix $R$ is unitary, $RR^{\dagger}=1$, with electron and hole submatrices,
\begin{equation}
R=\begin{pmatrix}
R_{\rm ee}&R_{\rm eh}\\
R_{\rm he}&R_{\rm hh}
\end{pmatrix},\label{RblockDef}
\end{equation}
describing normal reflection (from electron to electron or from hole to hole) and Andreev reflection (from electron to hole or vice versa). The linear response conductance (in the zero-temperature limit) is given by
\begin{equation}
G=G_{0}\,{\rm Tr}\,\bigl(1-R_{\rm ee}^{\vphantom{\dagger}}R_{\rm ee}^{\dagger}+R_{\rm he}^{\vphantom{\dagger}}R_{\rm he}^{\dagger}\bigr)=2G_{0}\,{\rm Tr}\,R_{\rm he}^{\vphantom{\dagger}}R_{\rm he}^{\dagger},\label{GBTK}
\end{equation}
with $G_{0}=e^{2}/h$ the conductance quantum. It it convenient to rewrite this without reference to the submatrices,
\begin{equation}
G/G_{0}=\tfrac{1}{2}\,{\rm Tr}\,\bigl(1-R^{\vphantom{\dagger}}\tau_{z}R^{\dagger}\tau_{z}\bigr),\;\;\tau_{z}=\begin{pmatrix}
1&0\\
0&-1
\end{pmatrix}.\label{Geh}
\end{equation}
The Pauli matrix $\tau_{z}$ acts on the electron and hole degrees of freedom.

The edge channels of a spin-triplet \textit{p}-wave superconductor are self-conjugate Majorana modes. It is then useful, following  Refs.\ \onlinecite{Ser10,Li12}, to transform from the electron-hole basis to the Majorana basis,
\begin{equation}
R\mapsto{\cal U} R{\cal U}^{\dagger},\;\;{\cal U}=\sqrt{\frac{1}{2}}\begin{pmatrix}
1&1\\
-i&i
\end{pmatrix}.\label{RUmapping}
\end{equation}
Electron-hole symmetry at the Fermi level requires that the scattering matrix elements are real in the Majorana basis, so $R^{\dagger}=R^{\rm T}=R^{-1}$. Because the Pauli matrix $\tau_{z}$ transforms into ${\cal U}\tau_{z}{\cal U}^{\dagger}=-\tau_{y}$, the conductance is given by
\begin{equation}
G/G_{0}=\tfrac{1}{2}\,{\rm Tr}\,\bigl(1-R^{\vphantom{\dagger}}\tau_{y}R^{\rm T}\tau_{y}\bigr),\;\;\tau_{y}=\begin{pmatrix}
0&-i\\
i&0
\end{pmatrix}.\label{GMajorana}
\end{equation}

In what follows we will work in the electron-hole basis \eqref{Geh} for spin-singlet \textit{d}-wave pairing (when the modes are not self-conjugate) and in the Majorana basis \eqref{GMajorana} for spin-triplet \textit{p}-wave pairing.

\section{Random-matrix theory}
\label{chiraledge}

The effect on the Andreev conductance of a topological phase transition at the Josephson junction can be analyzed in a model-independent way by means of random-matrix theory. We will first do this for an unpaired chiral Majorana mode and then for a pair of helical Majorana modes. The Josephson junction cannot undergo a topological phase transition for chiral Dirac modes, and will generically not for an even number of chiral Majorana modes, so these two cases are not considered in this section.

\subsection{Chiral Majorana mode}
\label{conductancesigma}

The conductance depends on the magnetic flux in a way which is restricted by electron-hole symmetry at the Fermi level. The restrictions are most severe for an unpaired chiral Majorana mode: The only phase shift allowed by electron-hole symmetry is $\pi$ (mod $2\pi$), so the conductance remains flux independent except when the enclosed flux is $h/4e$ (mod $h/2e$). Let us investigate this case in some detail.

A single Majorana mode corresponds to scalars $S_{\rm J}$ and $t_{\rm edge}$, to a row vector $t_{\rm edge,N}$, and to a column vector $t_{\rm N,edge}$. The contraction $t_{\rm edge,N}\tau_{y}t_{\rm edge,N}^{\rm T}$ produces a scalar, which vanishes because $\tau_{y}$ is an antisymmetric matrix. This eliminates one term when we substitute Eq.\ \eqref{Rdef} into Eq.\ \eqref{GMajorana}. What remains is
\begin{align}
G/G_{0}={}&\tfrac{1}{2}\,{\rm Tr}\,\bigl(1-r_{\rm N}\tau_{y}r_{\rm N}^{\rm T}\tau_{y}\bigr)\nonumber\\
&-\frac{S_{\rm J}}{1-S_{\rm J}t_{\rm edge}}t_{\rm edge,N}\tau_{y}r_{\rm N}^{\rm T}\tau_{y}t_{\rm N,edge}.\label{GSMajorana}
\end{align}

Because $S_{\rm J}$ is an orthogonal matrix consisting of a single matrix element, it can only equal $\pm 1$. Including a $\pi$ phase shift from the winding around the ring, we define
\begin{equation}
\sigma=-S_{\rm J}\in\{+1,-1\}\label{sigmaSJ}
\end{equation}
as the topological quantum number of the Josephson junction. The effective reflection matrix $R$ of the NS interface, constructed from Eq.\ \eqref{Rdef}, inherits this topological quantum number,
\begin{equation}
{\rm Det}\,R=\sigma.\label{DetRsigma}
\end{equation}
This follows from general considerations for a topological quantum number in symmetry class D,\cite{Akh11} but one can check it explicitly from Eq.\ \eqref{Rdef}.

The conductance for a ring with an unpaired Majorana mode is flux independent --- except at topological phase transitions, when $\sigma$ changes sign and the conductance jumps by an amount $\pm\delta G$ with
\begin{equation}
\delta G=G(\sigma=1)-G(\sigma=-1).\label{deltaGdef}
\end{equation}
Using Eq.\ \eqref{GSMajorana} this can be written as
\begin{equation}
\delta G/G_{0}=\frac{2}{1-t_{\rm edge}^2}t_{\rm edge,N}\tau_{y}r_{\rm N}^{\rm T}\tau_{y}t_{\rm N,edge}.\label{deltaGdefLoop}
\end{equation}

For a disordered NS interface we may consider an ensemble of scattering matrices $S$, generated by varying the disorder realization. A simple choice is the circular real ensemble (CRE) of class-D random-matrix theory,\cite{Ser10,Alt97,Bee11} for which $S$ is uniformly distributed in ${\rm SO}(2N+1)$: The group of orthogonal $(2N+1)\times (2N+1)$ matrices $O$ with ${\rm Det}\,O=1$. The integer $N$ counts the number of modes in the contact with the normal metal, including the spin degree of freedom. The factor of $2$ in $2N+1$ accounts for the electron-hole degree of freedom and the $+1$ refers to the unpaired Majorana mode. 

The effective reflection matrix $R$, constructed from $S$ via Eq.\ \eqref{Rdef}, inherits the uniform CRE distribution in ${\rm O}_{\sigma}(2N)$ --- the set of $2N\times 2N$ orthogonal matrices with determinant $\sigma$. The uniformity of $R\in{\rm O}_{\sigma}(2N)$ is a consequence of the uniformity of $S\in{\rm SO}(2N+1)$ because the transformation $S\mapsto S\cdot (U_{0}\oplus 1)$ with $U_{0}\in{\rm SO}(2N)$ maps  $R$ onto $R\cdot U_{0}$ without changing the determinant.

If $N=1$ or $N=2$ the distribution of $\delta G$ follows directly from the known\cite{Bee11} distribution $P_{\sigma}(G)$ of the conductance in the CRE: In both these cases $P_{-}(G)=\delta(G-2G_{0})$ is a delta-function distribution, so we may equate $P(\delta G)=P_{+}(G=\delta G+2G_{0})$. Since $P_{+}(G)=\delta(G)$ for $N=1$ and uniform in $[0,4G_{0}]$ for $N=2$, we arrive at
\begin{equation}
\begin{split}
&P(\delta G)=\delta(\delta G+2G_{0}),\;\;{\rm for}\;\;N=1,\\
&P(\delta G)=1/4G_{0},\;\;-2G_{0}\leq G\leq 2G_{0},\;\;{\rm for}\;\;N=2.
\end{split}\label{PGN12}
\end{equation}

\begin{figure}[tb]
\centerline{\includegraphics[width=0.8\linewidth]{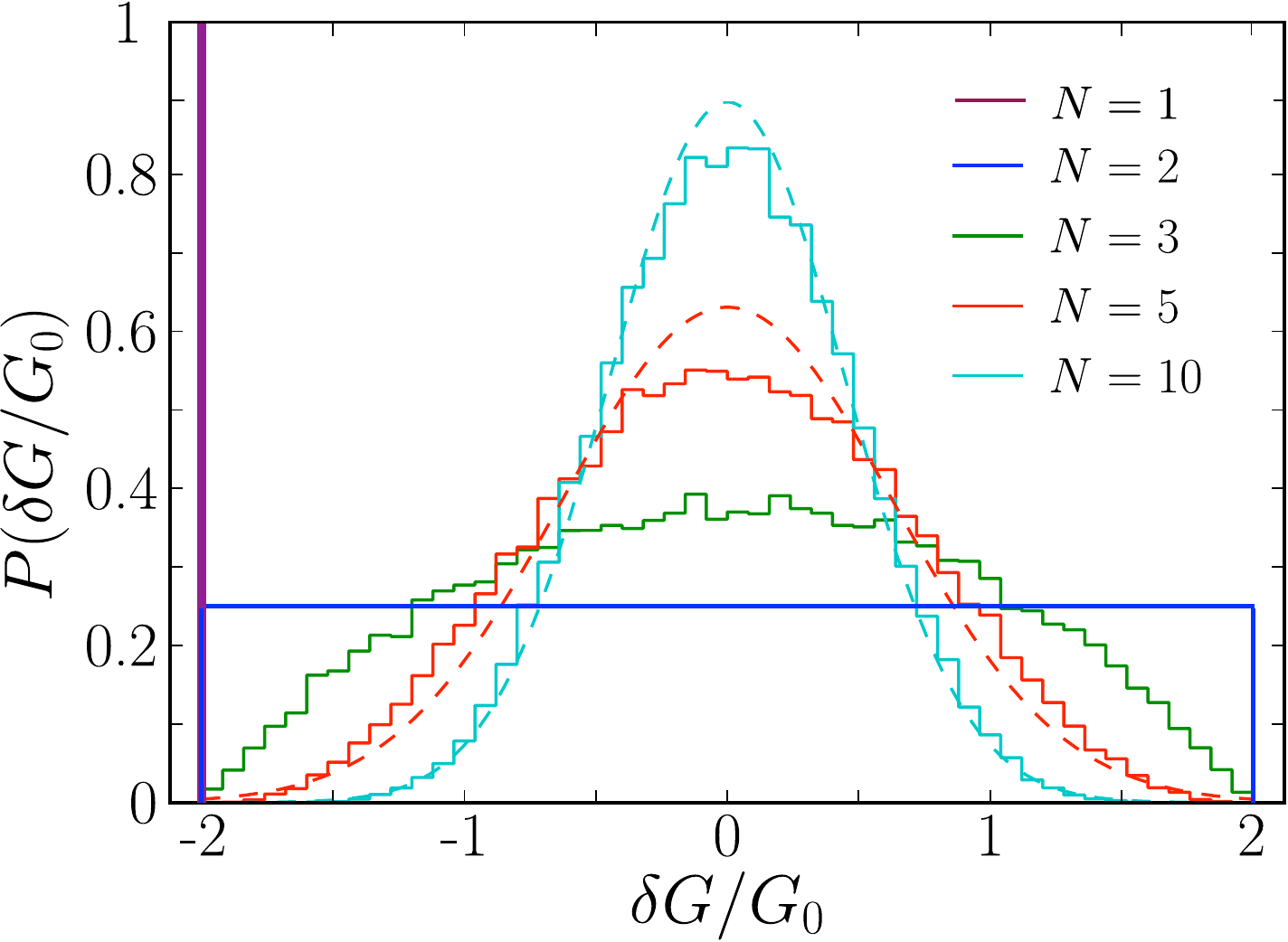}}
\caption{
Probability distribution of the change $\delta G$ in the conductance of the superconducting ring when the Josephson junction switches from topologically trivial to nontrivial. These are results from random-matrix theory in the circular real ensemble (CRE, symmetry class D), for an unpaired chiral Majorana mode circulating along the ring and $N$ modes in the contact to the normal metal. The solid lines for $N=1,2$ are from Eq.\ \eqref{PGN12}, the histograms are obtained numerically by averaging the scattering matrix $S$ uniformly over ${\rm SO}(2N+1)$, and the dashed curves are the large-$N$ Gaussian limit \eqref{VarGGaussian}.
}
\label{fig_RMTD}
\end{figure}

For $N>2$ the knowledge of $P_{\sigma}(G)$ is not sufficient to determine $P(\delta G)$, but it can be determined directly from the uniform distribution of $S$ in ${\rm SO}(2N+1)$. We have carried out this calculation numerically for small $N$, see Fig.\ \ref{fig_RMTD}.

For large $N$ we can approximate the matrix elements of $S$ by independent Gaussians, of zero mean and variance $1/2N$. We define the unit vectors
\begin{equation}
\hat{u}=\frac{it_{\rm edge,N}\tau_{y}}{(1-t_{\rm edge}^{2})^{1/2}},\;\;\hat{v}=\frac{i\tau_{y}t_{\rm N,edge}}{(1-t_{\rm edge}^{2})^{1/2}},\label{uvdef}
\end{equation}
so that the conductance change \eqref{deltaGdefLoop} is given by 
\begin{equation}
\delta G/G_{0}=-2\sum_{n,m=1}^{2N}\hat{u}_{n}\hat{v}_{m}(r_{\rm N})_{mn}.\label{uvdeltaG}
\end{equation}

In the large-$N$ Gaussian approximation, $\delta G/G_{0}$ is the sum of Gaussians with zero mean and variance $(2/N)(\hat{u}_{n}\hat{v}_{m})^{2}$, so its distribution is again a Gaussian with zero mean and variance
\begin{equation}
{\rm Var}\,(\delta G/G_{0})=\frac{2}{N}\sum_{n,m=1}^{2N}(\hat{u}_{n}\hat{v}_{m})^{2}=\frac{2}{N},\;\;N\gg 1.\label{VarGGaussian}
\end{equation}

\subsection{Helical Majorana modes}
\label{helicalMmode}

Spin-triplet pairing with time-reversal symmetry can produce a pair of counterpropagating (helical) Majorana modes. This is symmetry class DIII. In the Majorana basis the scattering matrix is orthogonal, as in class D, with the additional time-reversal symmetry condition\cite{Alt97}
\begin{equation}
S=\tau_{y}S^{\rm T}\tau_{y}.\label{SSTrelation}
\end{equation}
This is equivalent to the requirement that the matrix product $\tilde{S}\equiv i\tau_{y}S$ is both orthogonal ($\tilde{S}^{\dagger}=\tilde{S}^{\rm T}=\tilde{S}^{-1}$) and antisymmetric ($\tilde{S}^{\rm T}=-\tilde{S}$). The class-DIII random-matrix ensemble (T-CRE) is generated by drawing a matrix $O$ from the CRE and constructing
\begin{equation}
\tilde{S}\equiv i\tau_{y}S=O\cdot i\tau_{y}\cdot O^{\rm T},\;\;O\in{\rm SO}(4M+2).\label{calSSrelation}
\end{equation}
The channel number $M=N/2$ again only refers to the orbital degree of freedom, each mode having a twofold Kramers degeneracy. 

The Josephson junction breaks time-reversal symmetry, for $\phi\neq 0$ (mod $\pi$), so it may couple the two edge states and cause backscattering at the junction. In the simplest description (not made in the numerical calculations of the next section) we neglect this coupling and set
\begin{equation}
S_{\rm J}=-\sigma\begin{pmatrix}
1&0\\
0&1
\end{pmatrix}\Rightarrow \tilde{S}_{\rm J}\equiv i\tau_{y}S_{\rm J}=-i\sigma\tau_{y}.\label{SJcalS}
\end{equation}
The Pfaffian of $\tilde{S}_{\rm J}$ is the class-DIII topological invariant,\cite{Ful11}
\begin{equation}
\sigma=-{\rm Pf}\,\tilde{S}_{\rm J}\in\{+1,-1\}.\label{PfcalS}
\end{equation}

The effective reflection matrix $R$, constructed from $S$ and $S_{\rm J}$ via Eq.\ \eqref{Rdef}, inherits this topological invariant,
\begin{equation}
{\rm Pf}\,(i\tau_{y}R)=\sigma,
\end{equation}
and also inherits the uniform distribution of the T-CRE:
\begin{equation}
\tilde{R}\equiv i\tau_{y}R=O\cdot i\tau_{y}\cdot O^{\rm T},\;\;O\in{\rm O}_{\sigma}(4M).\label{ROrelation}
\end{equation}

We seek the probability distribution $P(\delta G)$ of the conductance change upon a topological phase transition in the T-CRE. For $N=2M=2$ the known\cite{Bee11} probability distribution $P_{\sigma}(G)$ of the conductance gives sufficient information, since $P_{-}(G)=\delta(G-4G_{0})\Rightarrow P(\delta G)=P_{+}(G=\delta G+4G_{0})$, resulting in
\begin{align}
P(\delta G/G_{0})={}&\frac{1}{8\sqrt{1+\delta G/4G_{0}}},\;\;-4\leq \delta G/G_{0}\leq 0,\nonumber\\
&{\rm for}\;\;N=2.\label{PGDIII}
\end{align}

\begin{figure}[tb]
\centerline{\includegraphics[width=0.8\linewidth]{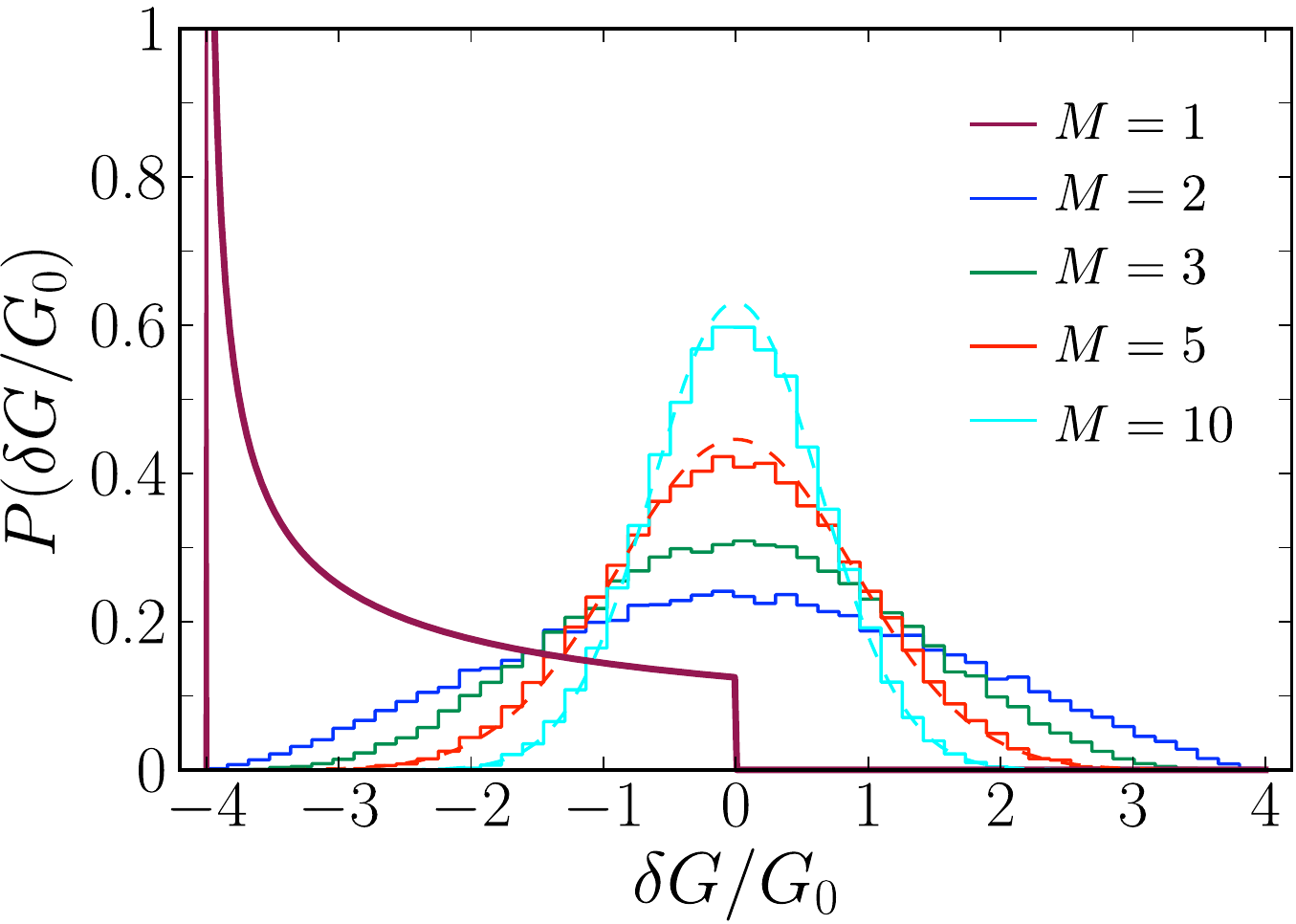}}
\caption{
Probability distribution of the conductance change upon a topological phase transition in symmetry class DIII, involving a pair of helical Majorana  modes circulating along the ring. The contact to the normal metal has $M$ orbital modes, each with a twofold Kramers degeneracy. The results are from the time-reversally invariant circular real ensemble (T-CRE) of random-matrix theory. The solid line for $M=1$ is from Eq.\ \eqref{PGDIII}, the histograms are obtained numerically, and the dashed curves are the large-$N$ Gaussian limit \eqref{VarGGaussianDIII}.
}
\label{fig_RMTDIII}
\end{figure}

The distribution $P(\delta G)$ for $M>1$ has been obtained by generating random matrices $O$ uniformly in ${\rm SO}(4M+2)$ and then constructing $S$ in the T-CRE via Eq.\ \eqref{calSSrelation}. Results are shown in Fig.\ \ref{fig_RMTDIII}. For $N\gg 1$ we have again a Gaussian distribution with zero mean and variance
\begin{equation}
{\rm Var}\,(\delta G/G_{0})=\frac{8}{N},\;\;N\gg 1,\label{VarGGaussianDIII}
\end{equation}
four times larger than Eq.\ \eqref{VarGGaussian} because of the twofold Kramers degeneracy.

\section{Results for model Hamiltonians}
\label{sec:models}

The analytical considerations of the previous section rely only on the fundamental symmetries of the Hamiltonian, without reference to a particular model. Here we present numerical results for model Hamiltonians in the various symmetry classes.

\subsection{Chiral pair potentials}
\label{chiralpairing}

We consider a two-dimensional superconductor in the $x$-$y$ plane, with pair potential $\hat\Delta$ dependent on the momentum $\bm{p}=-i\hbar(\partial_x,\partial_y)$. The Bogoliubov-De Gennes Hamiltonian, in the electron-hole basis, has the form
\begin{equation}
{\cal H}=\begin{pmatrix} 
	H & \hat\Delta\\
	\hat\Delta^{\dagger} & -H^*
	 \end{pmatrix}.\label{HBdG}
\end{equation}
It contains the single-particle Hamiltonian
\begin{equation}
H=(\bm{p} - e\bm{A})^2/2m_\text{eff}-\mu + U,\label{Hsingledef}
\end{equation}
with $\bm{A}(\bm{r})$ the vector potential, $m_\text{eff}$ the effective mass, $\mu$ the chemical potential, and $U(\bm{r})$ an electrostatic disorder potential. (We conveniently set the electronic charge to $+e$.)

The electron-hole symmetry relations are different for the different symmetry classes,
\begin{subequations}
\label{CDsymmetry}
\begin{align}
&\hat{\Delta}_{\rm D}^{\dagger}=-\hat{\Delta}_{\rm D}^{\ast}\Rightarrow{\cal H}_{\rm D}=-\tau_{x}{\cal H}_{\rm D}^{\ast}\tau_{x},\label{Dsymmetry}\\
&\hat{\Delta}_{\rm C}^{\dagger}=\hat{\Delta}_{\rm C}^{\ast}\Rightarrow{\cal H}_{\rm C}=-\tau_{y}{\cal H}_{\rm C}^{\ast}\tau_{y}.\label{Csymmetry}	
\end{align}
\end{subequations}
As specific models we take in class D the spin-triplet chiral \textit{p}-wave pairing
\begin{equation}
\hat\Delta_{\rm D}=p_{\rm F}^{-1}\{\Delta(\bm{r}),p_{x}+ip_{y}\},\label{DeltaD}
\end{equation}
with operators symmetrized by $\{a,b\}=\tfrac{1}{2}(ab+ba)$. In class C we take the spin-singlet chiral \textit{d}-wave pairing
\begin{subequations}
\label{DeltaC}
\begin{align}
&\hat\Delta_{\rm C}=\sum_{\alpha,\beta}(\bm{p}-e\bm{A})_{\alpha}M_{\alpha\beta}(\bm{p}+e\bm{A})_{\beta},\label{DeltaCa}\\
&\bm{M}(\bm{r})=p_{\rm F}^{-2}\Delta(\bm{r})\begin{pmatrix}
1&i\\
i&-1
\end{pmatrix}.\label{DeltaCb}
\end{align}
\end{subequations}
Both pair potentials properly produce a gauge invariant Bogoliubov-De Gennes Hamiltonian,\cite{note1}
\begin{equation}
e^{-i\chi\tau_{z}}{\cal H}(\bm{A},\Delta)e^{i\chi\tau_{z}}={\cal H}\left(\bm{A}-\frac{\hbar}{e}\nabla\chi,e^{-2i\chi}\Delta\right).\label{gaugeinvariance}
\end{equation}

Since $\hat{\Delta}_{\rm D}=\Delta_{0} e^{i\theta}$ and $\hat{\Delta}_{\rm C}=\Delta_{0} e^{2i\theta}$ when $\Delta(\bm{r})=\Delta_{0}$, $\bm{A}=0$, and $\bm{p}=p_{\rm F}(\cos\theta,\sin\theta)$, the magnitude of the gap is independent of the orientation. We expect that more general anisotropic models will give the same qualitative results --- provided that the gap does not vanish in any direction.

The ring has a weak link of length $R_{\rm outer}-R_{\rm inner}$, with $R_{\rm inner}$ and $R_{\rm outer}$ the inner and outer radius of the ring. We assume that the ring is wide compared to the London penetration depth $\lambda_{\rm L}$ but narrow compared to the Josephson penetration depth $\lambda_{\rm J}$,
\begin{equation}
\lambda_{\rm L}\ll R_{\rm outer}-R_{\rm inner}\ll\lambda_{\rm J}.\label{lambdaLJ}
\end{equation}
The first inequality ensures that the magnetic field is screened from the ring except at the weak link, along which a flux $\Phi$ can enter. The second inequality prevents vortices to appear inside the weak link. The gauge invariant phase difference across the weak link then has a uniform value $\phi=(2e/\hbar)\Phi$. We will use a gauge with a real uniform order parameter $\Delta(\bm{r})=\Delta_{0}$ and a delta-function vector potential
\begin{equation}
\bm{A}=\Phi\,\theta(-y)\delta(x)\hat{x},\label{Adef}
\end{equation}
for a Josephson junction at $x=0$ (aligned along the negative $y$-axis).

\subsection{Helical pair potential}
\label{helicalpairing}

We construct a model Hamiltonian with helical pairing from two time-reversed copies of the class-D chiral \textit{p}-wave pairing, $p_{x}\pm ip_{y}$. Spin-orbit coupling of the Rashba form couples the spin-up $p_{x}+ip_{y}$ sector with the spin-down $p_{x}-ip_{y}$ sector, promoting the symmetry class from D to DIII. 

The Bogoliubov-De Gennes Hamiltonian \eqref{HBdG} contains the single-particle Hamiltonian
\begin{equation}
H_{\rm DIII} = \bigl[(\bm{p} - e\bm{A})^2/2m_\text{eff}-\mu + U\bigr]\sigma_0 + \alpha_{\rm so}(p_x\sigma_y-p_y\sigma_x),\label{HsingleDIII}
\end{equation}
where $\sigma_x, \sigma_y, \sigma_z$ are the Pauli matrices acting on the spin degree of freedom and $\sigma_0$ is the corresponding unit matrix. The spin-orbit coupling strength is denoted by $\alpha_{\rm so}$. The helical pair potential is given by
\begin{equation}
\hat\Delta_{\rm DIII} = p_\text{F}^{-1}\{\Delta(\bm r), p_x\sigma_z+ip_y\sigma_0\}.
  \label{DeltaDIII}
\end{equation}

The electron-hole symmetry requirement in class DIII is the same as in class D, cf.\ Eq.\ \eqref{CDsymmetry},
\begin{equation}
\hat{\Delta}_{\rm DIII}^\dagger = -\hat{\Delta}_\text{DIII}^\ast \Rightarrow {\cal H}_{\rm DIII}=-\tau_x {\cal H}_{\rm DIII}^\ast\tau_x.
  \label{DIIIph}
\end{equation}
For $\bm{A}=0$ and real $\Delta$ the class-DIII Hamiltonian satisfies the time reversal symmetry
\begin{equation}
  {\cal H}_{\rm DIII} = \sigma_y {\cal H}_{\rm DIII}^\ast \sigma_y
  \label{DIIItimerevesal}.
\end{equation}

\subsection{Topological phase transition at the Josephson junction}
\label{topphaseJJ}

\begin{figure}[tb]
\centerline{\includegraphics[width=0.8\linewidth]{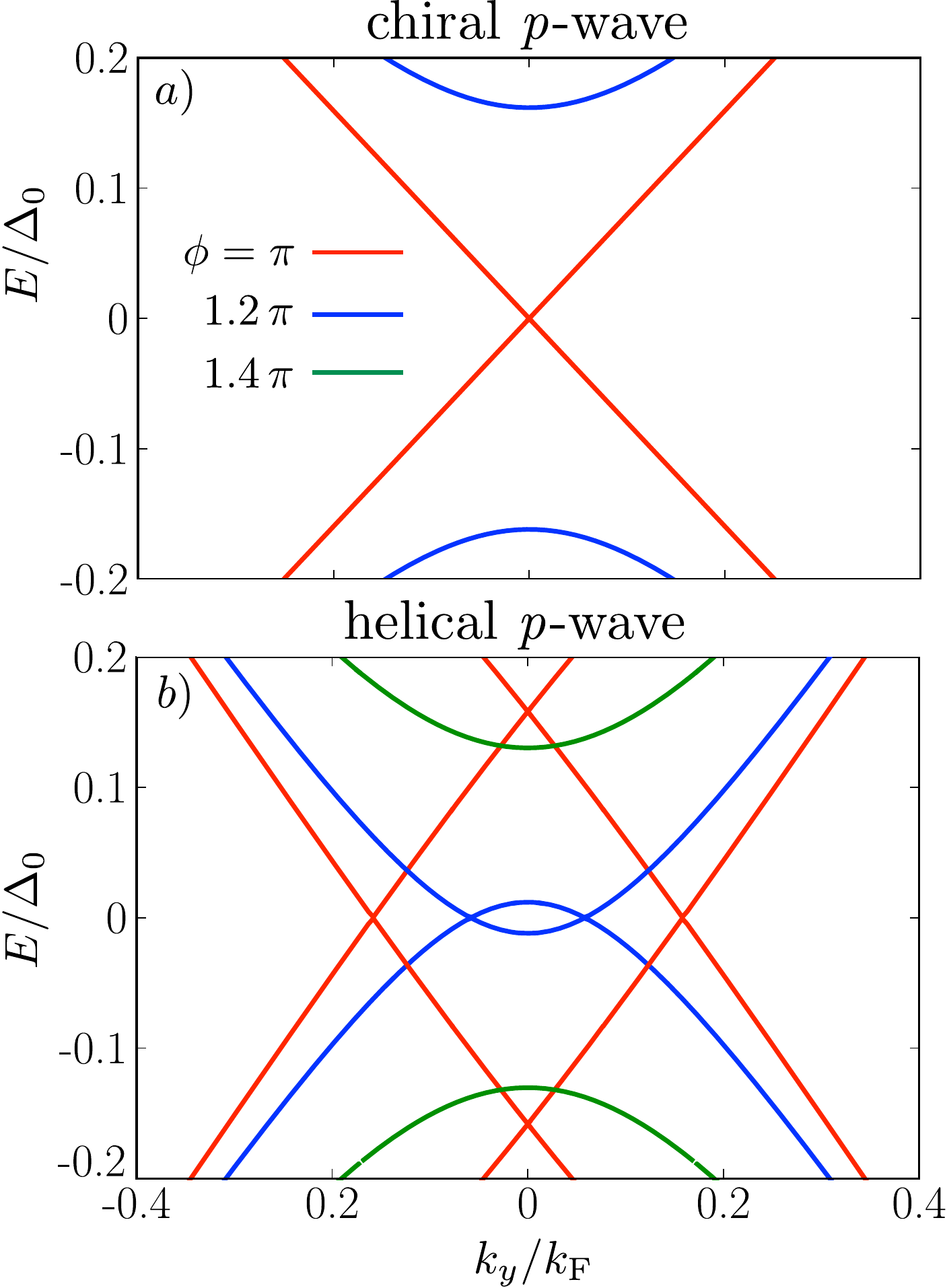}}
\caption{
Excitation spectrum of an infinitely long Josephson junction along the $y$-axis, for different values of the phase difference $\phi$, calculated numerically from the discretized Bogoliubov-De Gennes Hamiltonian \eqref{HBdG}.\cite{parameters} Panel a) is for the class-D chiral \textit{p}-wave pair potential \eqref{DeltaD} and panel b) for the class-DIII helical \textit{p}-wave pair potential \eqref{DIIIph}. The closing of the excitation gap at $\phi=\pi$ is topologically protected.
}
\label{fig_spectra}
\end{figure}

The phase transition in classes D and DIII is evidenced by a closing of the excitation gap at the Josephson junction when $\phi=\pi$ (mod $2\pi$). The gap closing and reopening is accompanied\cite{Akh11,Ful11} by a sign change of the topological quantum number $\sigma={\rm Det}\,R$ (in class D) or $\sigma={\rm Pf}\,i\tau_{y}R$ (in class DIII). In Fig.\ \ref{fig_spectra} we illustrate the gap closing for the chiral and helical \textit{p}-wave pairings \eqref{DeltaD} and \eqref{DIIIph}. 

Away from  $\phi=\pi$ the gap immediately opens in class D, while the gap closing persists for some range of $\phi$ in class DIII. This is a consequence of translational invariance along the weak link, see App.\ \ref{spatialsymmetry}. Only the gap closing at $\phi=\pi$ is topologically protected.

\begin{figure}[tb]
\centerline{\includegraphics[width=0.8\linewidth]{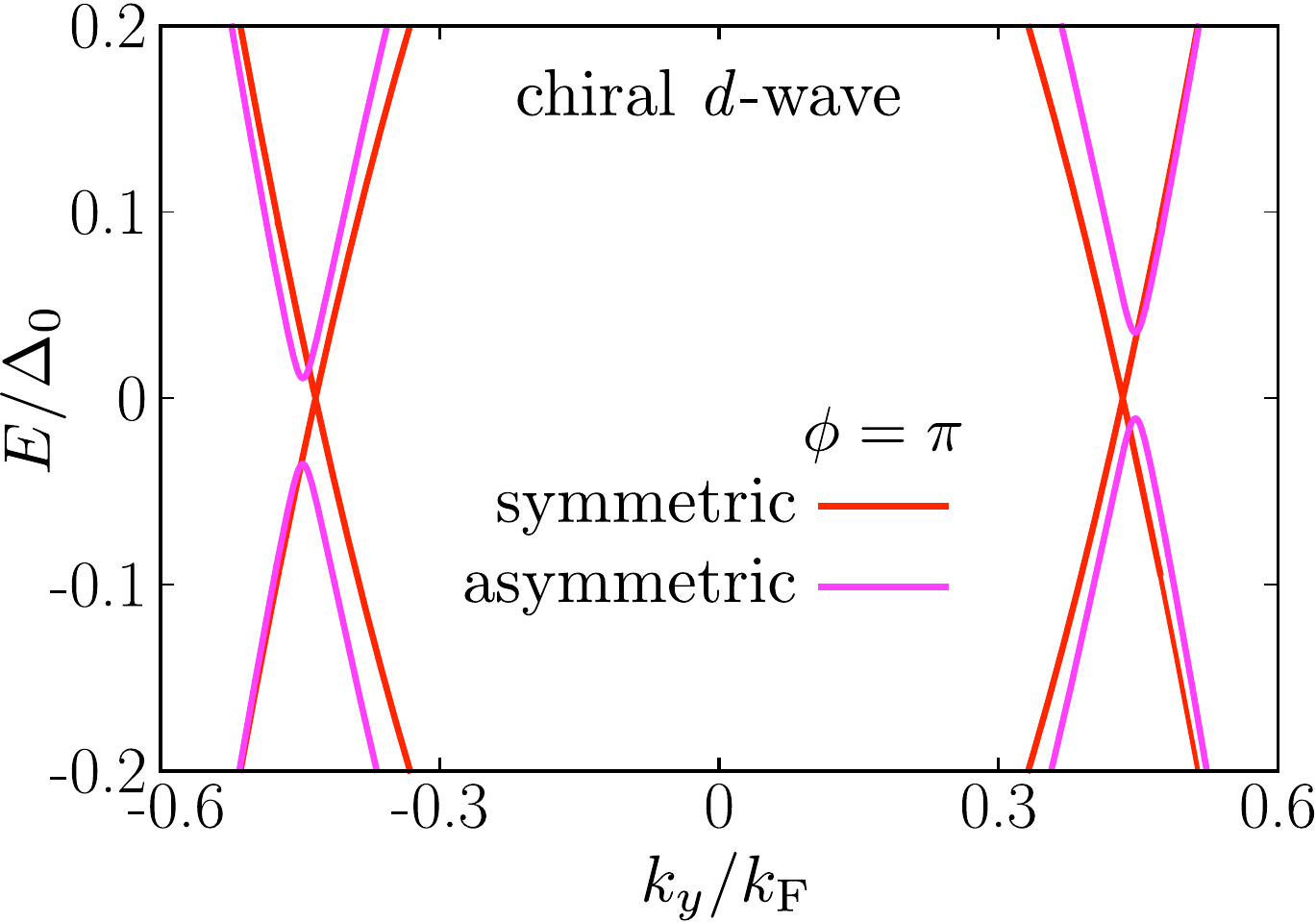}}
\caption{
Same as Fig.\ \ref{fig_spectra}, for a class-C chiral \textit{d}-wave pair potential. The two curves are for $\phi=\pi$, with and without $\pm x$ symmetry of the Josephson junction along the $y$-axis. The gap closing now has no topological protection, but requires a spatial symmetry.
}
\label{fig_spectrumC}
\end{figure}

In class C there is no gap closing that is \textit{topologically} protected. However, as explained in App.\ \ref{spatialsymmetry}, the combination of translational invariance along the $y$-axis and $x\mapsto -x$ reflection symmetry allow for a gap closing at $\phi=\pi$ (mod $2\pi$). We show this in Fig.\ \ref{fig_spectrumC} for the chiral \textit{d}-wave pairing \eqref{DeltaC}. Disorder will break these symmetries and remove the gap closing.

\subsection{Numerical results}
\label{numericsim}

\begin{figure*}[tb]
\centerline{\includegraphics[width=1\linewidth]{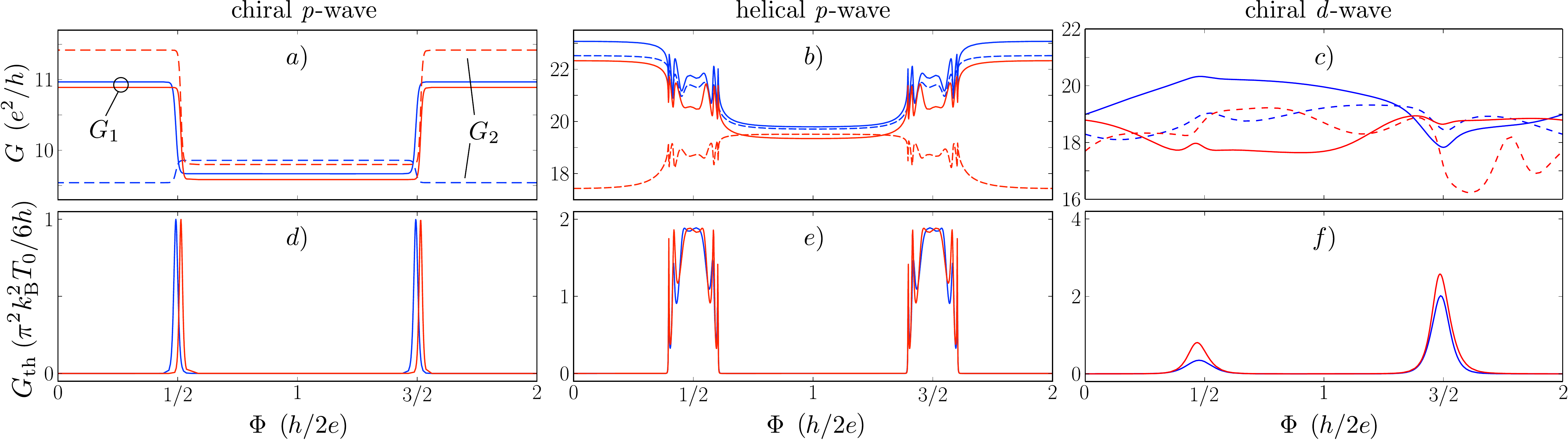}}
\caption{
Electrical conductance (upper row) and thermal conductance (lower row) in the ring geometry of Fig.\ \ref{fig:setup}, as a function of the enclosed flux $\Phi$, for the three pair potentials \eqref{DeltaD}, \eqref{DIIIph}, and \eqref{DeltaC}. The red and blue curves are calculated numerically for two different realizations of the disorder potential.\cite{parameters} The thermal conductance is measured between the inner and outer perimeter of the ring. The electrical conductance is measured either between the inner perimeter and ground ($G_1$, solid curves) or between the outer perimeter and ground ($G_2$, dashed curves).
}
\label{fig:results}
\end{figure*}

We have discretized the Bogoliubov-De Gennes Hamiltonian on a square lattice (lattice constant $a$, hopping amplitude $t$, see App.\ \ref{app:discretization}). The geometry is shown in Fig.\ \ref{fig:setup}, with a pair of normal-metal leads (width $30\,a$) attached to the inner and outer perimeter (radii $50\,a$ and $100\,a$). 

The leads are modeled by setting $\Delta_{0}$, $\bm{A}$, $\alpha_{\rm so}$, and $U$ all equal to zero, at a chemical potential $\mu_{\rm N}=0.5\,t$ for which there are $M=6$ orbital modes. Each of these modes is spin-degenerate when coupled to the chiral \textit{d}-wave or helical \textit{p}-wave superconductor, and nondegenerate when coupled to the chiral \textit{p}-wave superconductor.

At the weak link the hopping matrix elements are reduced such that the transmission probability per mode is $\lesssim 0.1$. Disorder in the superconductor is introduced by a random on-site potential $U(x,y)$, uniformly distributed in the interval $(-U_{\rm disorder}/2, U_{\rm disorder}/2)$. We took $U_{\rm disorder}=0.7\,t$.

We solve the scattering problem numerically,\cite{Wim09} to obtain the scattering matrix ${\cal S}$ for the electron (e) and hole (h) modes incident on the superconductor from the inner contact (1) and the outer contact (2). We then calculate the Andreev conductances $G_{1}$ and $G_{2}$ from the outer or inner contact to ground,
\begin{align}
&G_{1}=\frac{e^{2}}{h}\,{\rm Tr}\,\bigl(1-{\cal S}_{1e,1e}^{\vphantom{\dagger}}{\cal S}_{1e,1e}^{\dagger}+{\cal S}_{1h,1e}^{\vphantom{\dagger}}{\cal S}_{1h,1e}^{\dagger}\bigr),\label{G1def}\\
&G_{2}=\frac{e^{2}}{h}\,{\rm Tr}\,\bigl(1-{\cal S}_{2e,2e}^{\vphantom{\dagger}}{\cal S}_{2e,2e}^{\dagger}+{\cal S}_{2h,2e}^{\vphantom{\dagger}}{\cal S}_{2h,2e}^{\dagger}\bigr).\label{G2def}
\end{align}
This expression holds also at the gap closing, when there is a nonzero transmission probability between contacts 1 and 2, under the assumption that contact 2 is grounded for the measurement of $G_{1}$ and contact 1 is grounded for the measurement of $G_{2}$.

To probe the gap closing we also calculate the thermal conductance
\begin{align}
G_{\rm th}={}&\frac{\pi^{2}k_{\rm B}^{2}T_{0}}{6h}\,{\rm Tr}\,\bigl({\cal S}_{2e,1e}^{\vphantom{\dagger}}{\cal S}_{2e,1e}^{\dagger}+{\cal S}_{2h,1e}^{\vphantom{\dagger}}{\cal S}_{2h,1e}^{\dagger}\nonumber\\
&+{\cal S}_{2h,1h}^{\vphantom{\dagger}}{\cal S}_{2h,1h}^{\dagger}+{\cal S}_{2e,1h}^{\vphantom{\dagger}}{\cal S}_{2e,1h}^{\dagger}\bigr),\label{gthcalS}
\end{align}
which measures the thermal current between the inner and outer perimeter at temperatures $T_{0}$ and $T_{0}+\delta T$.

Results are shown in Fig.\ \ref{fig:results} for several disorder realizations. The results for the electrical conductance (top row) will be discussed in the next section, in connection with experimental probes for Dirac or Majorana edge modes. 

The thermal conductance (bottom row) is not easily measured, but is included here because it illustrates in a striking way the significance of topological protection for a gap closing. The change in sign of the topological quantum number at the class-D phase transition results in a peak of the thermal conductance that is quantized\cite{Akh11} in units of the thermal quantum $\pi^{2}k_{\rm B}^{2}T_{0}/6h$, see Fig.\ \ref{fig:results}d. The gap closing in class C has no topological protection, there is no sign change of a topological quantum number and no quantized peak, see Fig.\ \ref{fig:results}f. The class-DIII gap closing has topological protection (no backscattering) if it happens when the flux is a multiple of $h/4e$, so time-reversal symmetry is preserved. Disorder leads to small displacements of the transition away from $h/4e$, allowing for backscattering and resulting in a small deviation of the thermal conductance peak from the quantized value (see Fig.\ \ref{fig:results}e).

\section{Discussion}
\label{discuss}

The numerical results of Fig.\ \ref{fig:results}a,b,c illustrate how circulating edge states manifest themselves in the magnetoconductance of the ring. All three types of edge states introduce a flux dependence with a period of \textit{twice} the superconducting flux quantum $\Phi_{0}=h/2e$. The magnetoconductance oscillations are sample specific, depending on the disorder realization. The inner and outer perimeter experience a different impurity potential and thus show a different magnetoconductance, but with the same $h/e$ periodicity.  A measurement of the fundamental frequency of the Fourier transformed magnetoconductance would be an unambiguous way to establish the presence of circulating edge states.

The magnetoconductance contains additional information, it can identify unpaired (chiral or helical) Majorana modes. These produce jumps $\delta G$ in the conductance when the flux is close to an odd multiple of $\Phi_{0}/2$, associated with a topological phase transition at the Josephson junction. Both the sign and magnitude of $\delta G$ is disorder dependent and different at the inner and outer perimeter, but the flux $\Phi_{\rm c}$ at which the conductance jumps lines up. Notice that even when different disorder realizations cause a small shift in $\Phi_{\rm c}$ (compare red and blue curves in Fig.\ \ref{fig:results}a), the conductance at the inner and outer perimeter jumps at precisely the same $\Phi_{\rm c}$ (compare solid and dashed curves). This phase locking is a striking signature of a topological phase transition at the Josephson junction.

A measurement of the fundamental frequency component $\cos(\Phi/2\Phi_{0}+\alpha)$ of the magnetoconductance at the inner and outer perimeter of the ring would therefore show a random and uncorrelated phase $\alpha$ for Dirac modes, and a correlated phase peaked at $0$ (modulo $\pi$) for an unpaired Majorana mode.

These magnetoconductance signatures of Dirac and Majorana edge states can be helpful in the ongoing search for topological superconductors. Recent attention has focused on hybrid structures combining strong spin-orbit coupling with induced \textit{s}-wave superconductivity, to produce an effective chiral \textit{p}-wave pairing.\cite{Sac11,Wil12,Vel12} A superconducting ring deposited on a three-dimensional topological insulator would need a magnetic barrier along the perimeter to confine the edge states.\cite{Fu08,Law09} Alternatively, one might induce superconductivity in the two-dimensional electron gas of a semiconductor heterostructure with strong spin-orbit coupling,\cite{Sau10,Ali10} such as an InAs quantum well, and confine the edge states electrostatically by gate electrodes.

\acknowledgments

The numerical calculations were performed using the {\sc kwant} package developed by A. R. Akhmerov, C. W. Groth, X. Waintal, and M. Wimmer. This project was supported by the Dutch Science Foundation NWO/FOM, by an ERC Advanced Investigator Grant, and by a Lawrence Golub Fellowship.

\appendix

\section{Gap closings due to spatial symmetries}
\label{spatialsymmetry}

In symmetry classes D and DIII the closing of the excitation gap at the topological phase transition of the Josephson junction is topologically protected, meaning that disorder cannot open up the gap. However, in Fig.\ \ref{fig_spectrumC} we see a gap closing in symmetry class C, where the Josephson junction remains topologically trivial. Morover, in Fig.\ \ref{fig_spectra}b we see that the gap closing in the helical $p$-wave junction persists over a range of $\phi$, rather than being limited to a single $\phi$ as it is in class D. Both these features are due to spatial symmetries, as we now explain.

Translational symmetry along the weak link (the $y$-axis) permits us to consider the parallel momentum $k_{y}\equiv q$ as an external parameter. The Hamiltonian ${\cal H}(q)$ describes a zero-dimensional system which can undergo a topological phase transition as a function of the parameter $q$ in symmetry classes D and BDI.\cite{Ryu10} At this transition a $\mathbb{Z}_{2}$ topological quantum number changes sign, so to open up the gap requires, either, the breaking of a symmetry, or the merging of a pair of gap closings at a single value of $q$.

So how do we arrive in class D or BDI when we start out from class DIII or class C? As pointed out in Ref.\ \onlinecite{Dah12} in a different context, spatial symmetries can do this. 

Let us first show that ${\cal H}(q)$ is in class D for helical \textit{p}-wave pairing. On the one hand, the electron-hole symmetry relation \eqref{DIIIph} gives
\begin{equation}
{\cal H}(q)=-\tau_x {\cal H}^\ast(-q)\tau_x,\label{HqDIIIph}
\end{equation}
on the other hand, the helical \textit{p}-wave pairing has the additional symmetry
\begin{equation}
(\tau_z\otimes\sigma_y) {\cal H}(q) (\tau_z\otimes\sigma_y) = {\cal H}(-q).\label{DIIIextra}
\end{equation}
Taking these two equations together we arrive at a symmetry relation for ${\cal H}(q)$ at one single value of $q$,
\begin{equation}
\Omega {\cal H}(q)=-{\cal H}(q)\Omega,\;\; \Omega=(\tau_y\otimes\sigma_y){\cal K},\label{Omega}
\end{equation}
with ${\cal K}$ the operator of complex conjugation. Because $\Omega$ is an anti-unitary operator that squares to $+1$, this places ${\cal H}(q)$ in symmetry class D, with a topologically protected gap closing. Indeed, as we see in Fig.\ \ref{fig_spectra}b, the pair of gap closings at $\phi=\pi$ persist as $\phi$ is increased, until the gaps merge at $q=0$.

Turning now to class C, we will show that ${\cal H}(q)$ is in class BDI for chiral \textit{d}-wave pairing at $\phi=\pi$ and if the electrostatic potential $U(x)$ is $\pm x$ symmetric. Firstly, the class-C electron-hole symmetry relation reads
\begin{equation}
{\cal H}(q)=-\tau_y {\cal H}^\ast(-q)\tau_y.\label{HqCph}
\end{equation}
Secondly, for $\phi=\pi$ and $\bm{A}=0$ the Hamiltonian is real,
\begin{equation}
{\cal H}^{\ast}(q)={\cal H}(q).\label{HqrealC}
\end{equation}
Thirdly, the combination of $U(x)=U(-x)$ and $\Delta(x)=-\Delta(-x)$ at $\phi=\pi$ gives
\begin{equation}
\tau_{z}{\cal P}{\cal H}(q)={\cal H}(-q)\tau_{z}{\cal P},\label{calPH}
\end{equation}
where ${\cal P}$ is the reflection operator ($x\mapsto -x$). Eqs.\ \eqref{HqCph} and \eqref{calPH} together give
\begin{equation}
\Omega'{\cal H}(q)=-{\cal H}(q)\Omega',\;\;\Omega'=\tau_{x}{\cal P}{\cal K}.\label{Omegaprime}
\end{equation}
The anti-unitary operator $\Omega'$ also squares to $+1$. The symmetries \eqref{HqrealC} and \eqref{Omegaprime} place ${\cal H}(q)$ in class BDI, provided that $\phi=\pi$ and the reflection symmetry is unbroken. This is consistent with what is seen in Fig.\ \ref{fig_spectrumC}: The gap closing for chiral \textit{d}-wave pairing can be removed either by increasing $\phi$ away from $\pi$ or by breaking the $\pm x$ symmetry of the weak link.

\section{Gauge invariant discretization of the Bogoliubov-De Gennes Hamiltonian}
\label{app:discretization}

The discretization of the Bogoliubov-De Gennes Hamiltonian \eqref{HBdG} with a momentum dependent pair potential requires special care to ensure that the resulting tight-binding model is gauge invariant. We go through the steps in this Appendix. Following the established procedure of minimal coupling, we first discretize without a vector potential, then perform a gauge transformation on the lattice, and finally replace the gradient of the gauge field by the vector potential.

The discretization for $\bm{A}=0$ is carried out by replacing the differential operators by symmetric finite differences,
\begin{equation}
\partial_{x}f(x)\mapsto\frac{1}{2a}[f(x+a)-f(x-a)],
\end{equation}
to arrive at the tight-binding Hamiltonian
\begin{equation}
  t(n,m) =
  \begin{pmatrix}
	 t_{\rm ee}(n,m) & t_{\rm eh}(n,m) \\
	 t_{\rm he}(n,m) & t_{\rm hh}(n,m)
  \end{pmatrix}.
  \label{eq:tbmatrix}
\end{equation}
The indices $n, m$ label sites $\bm{r}_{n},\bm{r}_{m}$ of a square lattice (lattice constant $a$). The diagonal elements $n=m$ are the on-site energies and the off-diagonal elements $n\neq m$ are the hopping amplitudes between sites $n$ and $m$.

The single-particle kinetic energy gives the electron-electron matrix elements,
\begin{align}
&  t_{\rm ee}(n,n) = 4t -\mu + U(\bm{r}_{n}), \quad t=\hbar^2/2m_{\rm eff} a^2
  \label{eq:onsite-ee},\\
&  t_{\rm ee}(n,m\neq n) = 
  \begin{cases}
	 -t & \mbox{for $n,m$ nearest neighbours,} \\
	 0 & \mbox{otherwise,}
  \end{cases}\nonumber
\end{align}
and the hole-hole matrix elements $t_{\rm hh}(n,m) = -t_{\rm ee}(n,m)$.

For the chiral \textit{d}-wave pair potential \eqref{DeltaC}, still at $\bm{A}=0$, we obtain the nearest-neighbor hopping amplitudes
\begin{equation}
\begin{split}
 & t_{\rm eh}(n\pm a\hat x, n) = -\frac{1}{2q^2}\left[\Delta(\bm{r}_n)+\Delta(\bm{r}_n\pm a\hat x)\right],\\
 & t_{\rm eh}(n\pm a\hat y, n) = \frac{1}{2q^2}\left[\Delta(\bm{r}_n)+\Delta(\bm{r}_n\pm a\hat y)\right],
\end{split}
 \label{eq:the-nn}
\end{equation}
the next-nearest-neighbor hopping amplitudes
\begin{equation}
\begin{split}
&t_{\rm eh}(n+a\hat x\pm a\hat y, n) = \frac{\mp i}{4q^2}\left[\Delta(\bm{r}_n+a\hat x)+\Delta(\bm{r}_n\pm a\hat y)\right] ,\\
&t_{\rm eh}(n-a\hat x\pm a\hat y, n) = \frac{\pm i}{4q^2}\left[\Delta(\bm{r}_n-a\hat x)+\Delta(\bm{r}_n\pm a\hat y)\right] ,
\end{split}
\label{eq:the-nnn}
\end{equation}
and the on-site matrix elements
\begin{equation}
\begin{split}
  t_{\rm eh}(n,n){} ={}&  \frac{1}{2q^2}[\Delta(\bm{r}_n+a\hat x) + \Delta(\bm{r}_n-a\hat x) \\
    &- \Delta(\bm{r}_n+a\hat y) - \Delta(\bm{r}_n-a\hat y)].
    \end{split}
  \label{eq:onsitehe}
\end{equation}
(We have defined $q=k_{\rm F}a$.) These are all hopping amplitudes from hole to electron. The hopping amplitudes from electron to hole are related by Hermiticity,
\begin{equation}
t_{\rm he}(n,m)=t_{\rm eh}^{\ast}(m,n).
\end{equation}

We now introduce the vector potential $\bm{A}(\bm{r})=-(\hbar/e)\nabla\chi(\bm{r})$ by means of the gauge transformation
\begin{equation}
\begin{split}
&\tilde{t}(n,m)=e^{-i\tau_z\chi(\bm{r}_{n})}t(n,m)e^{i\tau_z\chi(\bm{r}_{m})},\\
&\tilde{\Delta}(\bm{r}_{n})=e^{-2i\chi(\bm{r}_{n})}\Delta(\bm{r}_{n}) .
\end{split}
\end{equation}
This is the lattice analogue of Eq.\ \eqref{gaugeinvariance}.

The effect on the electron-electron and hole-hole hopping amplitudes is the Peierls substitution,\cite{Pei33} 
\begin{equation}
\begin{split}
&\tilde{t}_{\rm ee}(n,m)=t_{\rm ee}(n,m)\exp\left(i\frac{e}{\hbar}\int_{m}^{n} \bm{A}\cdot d\bm{l}\right),\\
&\tilde{t}_{\rm hh}(n,m)=t_{\rm hh}(n,m)\exp\left(-i\frac{e}{\hbar}\int_{m}^{n} \bm{A}\cdot d\bm{l}\right).
\end{split}
\end{equation}
The line integral of the vector potential is taken along the lattice bond from site $m$ to site $n$, and with this prescription the Peierls substitution can also be applied to vector potentials that do not derive from a gauge field.

\begin{widetext}
The transformed electron-hole matrix hopping amplitudes for chiral \textit{d}-wave pairing are given by
\begin{equation}
\begin{split}
  \tilde{t}_{\rm eh}(n\pm a\hat x, n) &= \frac{-1}{2q^2}\left[e^{i \frac{e}{\hbar} \int_n^{n\pm a\hat x} \bm A\cdot{\rm d}\bm l}\tilde{\Delta}(\bm{r}_{n})+\tilde{\Delta}(\bm{r}_{n}\pm a\hat x)e^{-i \frac{e}{\hbar} \int_n^{n\pm a\hat x} \bm A\cdot{\rm d}\bm l}\right], \\
  \tilde{t}_{\rm eh}(n\pm a\hat y, n) &= \frac{1}{2q^2}\left[e^{i \frac{e}{\hbar} \int_n^{n\pm a\hat y} \bm A\cdot{\rm d}\bm l}\tilde{\Delta}(\bm{r}_{n})+\tilde{\Delta}(\bm{r}_{n}\pm a\hat y)e^{-i \frac{e}{\hbar} \int_n^{n\pm a\hat y} \bm A\cdot{\rm d}\bm l}\right],
  \end{split}
  \label{eq:the-nn-inv}
\end{equation}
\begin{equation}
\begin{split}
  \tilde{t}_{\rm eh}(n+a\hat x\pm a\hat y, n) &= \frac{\mp i}{4q^2}\left[e^{i \frac{e}{\hbar} \int_{n+a\hat x}^{n+a\hat x\pm a\hat y} \bm A\cdot{\rm d}\bm l}\tilde{\Delta}(\bm{r}_{n}+a\hat x)e^{-i \frac{e}{\hbar} \int_n^{n+a\hat x} \bm A\cdot{\rm d}\bm l}
  + e^{i \frac{e}{\hbar} \int_{n\pm a\hat y}^{n+a\hat x\pm a\hat y} \bm A\cdot{\rm d}\bm l}\tilde{\Delta}(\bm{r}_{n}\pm a\hat y)e^{-i \frac{e}{\hbar} \int_n^{n\pm a\hat y} \bm A\cdot{\rm d}\bm l}\right] ,\\
  \tilde{t}_{\rm eh}(n-a\hat x\pm a\hat y, n) &= \frac{\pm i}{4q^2}\left[e^{i \frac{e}{\hbar} \int_{n-a\hat x}^{n-a\hat x\pm a\hat y} \bm A\cdot{\rm d}\bm l}\tilde{\Delta}(\bm{r}_{n}-a\hat x)e^{-i \frac{e}{\hbar} \int_n^{n-a\hat x} \bm A\cdot{\rm d}\bm l}
  + e^{i \frac{e}{\hbar} \int_{n\pm a\hat y}^{n-a\hat x\pm a\hat y} \bm A\cdot{\rm d}\bm l}\tilde{\Delta}(\bm{r}_{n}\pm a\hat y)e^{-i \frac{e}{\hbar} \int_n^{n\pm a\hat y} \bm A\cdot{\rm d}\bm l}\right] ,
  \end{split}
  \label{eq:the-nnn-inv}
\end{equation}
\begin{equation}
\begin{split}
 \tilde{t}_{\rm eh}(n,n)=
  \frac{1}{2q^2}\biggl[ &\tilde{\Delta}(\bm{r}_{n}+a\hat x)e^{-i \frac{2e}{\hbar} \int_n^{n+a\hat x} \bm A\cdot{\rm d}\bm l} 
  + \tilde{\Delta}(\bm{r}_{n}-a\hat x)e^{-i \frac{2e}{\hbar} \int_n^{n-a\hat x} \bm A\cdot{\rm d}\bm l} 
     \\ 
    - & \tilde{\Delta}(\bm{r}_{n}+a\hat y)e^{-i \frac{2e}{\hbar} \int_n^{n+a\hat y} \bm A\cdot{\rm d}\bm l}
	- \tilde{\Delta}(\bm{r}_{n}-a\hat y)e^{-i \frac{2e}{\hbar} \int_n^{n-a\hat y} \bm A\cdot{\rm d}\bm l}\biggr].
	\end{split}
  \label{eq:onsitehe-inv}
\end{equation}

A similar calculation for the chiral \textit{p}-wave pairing \eqref{DeltaD} gives the electron-hole hopping amplitudes
\begin{equation}
\begin{split}
  \tilde{t}_{\rm eh}(n\pm a\hat x, n) &= \frac{\mp i}{4q}\left[e^{i \frac{e}{\hbar} \int_n^{n\pm a\hat x} \bm A\cdot{\rm d}\bm l}\tilde{\Delta}(\bm{r}_{n})+\tilde{\Delta}(\bm{r}_{n}\pm a\hat x)e^{-i \frac{e}{\hbar} \int_n^{n\pm a\hat x} \bm A\cdot{\rm d}\bm l}\right], \\
  \tilde{t}_{\rm eh}(n\pm a\hat y, n) &= \frac{\pm 1}{4q}\left[e^{i \frac{e}{\hbar} \int_n^{n\pm a\hat y} \bm A\cdot{\rm d}\bm l}\tilde{\Delta}(\bm{r}_{n})+\tilde{\Delta}(\bm{r}_{n}\pm a\hat y)e^{-i \frac{e}{\hbar} \int_n^{n\pm a\hat y} \bm A\cdot{\rm d}\bm l}\right].
  \end{split}
  \label{eq:the-inv}
\end{equation}
\end{widetext}
There are neither next-nearest-neighbor hoppings, nor on-site electron-hole matrix elements in this case. 

Notice that the discretized \textit{p}-wave pair potential \eqref{eq:the-inv} depends explicitly on the vector potential, while in the continuum representation \eqref{DeltaD} the vector potential enters only implicitly through $\Delta(\bm{r})$. All of this is required by gauge invariance.

Finally, we give the corresponding expressions for the helical \textit{p}-wave pairing \eqref{DeltaDIII}. There is now a spin degree of freedom $\sigma=\uparrow,\downarrow$, and the pair potential is diagonal in that index. The electron-hole hopping amplitudes are given by
\begin{equation}
\begin{split}
 & \tilde t_{\rm e\uparrow, h\uparrow}(n\pm a\hat x, n) = -\tilde t_{\rm e\downarrow,h\downarrow}(n\pm a\hat x, n) \equiv \tilde t_{\rm eh}(n\pm a\hat x,n) ,\\
  &\tilde t_{\rm e\uparrow, h\uparrow}(n\pm a\hat y, n) = \tilde t_{\rm e\downarrow,h\downarrow}(n\pm a\hat y, n) \equiv \tilde t_{\rm eh}(n\pm a\hat y,n),
    \end{split}
  \label{eq:DIIIts}
\end{equation}
where matrix elements without spin indices should be taken from Eq.\ \eqref{eq:the-inv}.


\begin{thebibliography}{99}
\bibitem{Vol97} G. E. Volovik, JETP Lett. \textbf{66}, 522 (1997).
\bibitem{Sen99} T. Senthil, J. B. Marston, and M. P. A. Fisher, Phys. Rev. B \textbf{60}, 4245 (1999).
\bibitem{Rea00} N. Read and D. Green, Phys. Rev. B \textbf{61}, 10267 (2000).
\bibitem{Has10} M. Z. Hasan and C. L. Kane, Rev. Mod. Phys. \textbf{82}, 3045 (2010).
\bibitem{Qi11} X.-L. Qi and S.-C. Zhang, Rev. Mod. Phys. \textbf{83}, 1057 (2011).
\bibitem{Kas00} S. Kashiwaya and Y. Tanaka, Rep. Prog. Phys. \textbf{63}, 1641 (2000).
\bibitem{Mac03} A. P. Mackenzie and Y. Maeno, Rev. Mod. Phys. \textbf{75}, 657 (2003).
\bibitem{Nan12} R. Nandkishore, L. Levitov, and A. Chubukov, Nature Phys. \textbf{8}, 158 (2012).
\bibitem{Fu08} L. Fu and C. L. Kane, Phys. Rev. Lett. \textbf{100}, 096407 (2008).
\bibitem{Ryu10} S. Ryu, A. Schnyder, A. Furusaki, and A. Ludwig, New J. Phys. \textbf{12}, 065010 (2010).
\bibitem{Qi09} X.-L. Qi, T. L. Hughes, S. Raghu, and S.-C. Zhang, Phys. Rev. Lett. \textbf{102}, 187001 (2009).
\bibitem{Tan09} Y. Tanaka, T. Yokoyama, A. V. Balatsky, and N. Nagaosa, Phys. Rev. B \textbf{79}, 060505 (2009).
\bibitem{Sat09} M. Sato and S. Fujimoto, Phys. Rev. B \textbf{79}, 094504 (2009).
\bibitem{Law09} K. T. Law, P. A. Lee, and T. K. Ng, Phys. Rev. Lett. \textbf{103}, 237001 (2009).
\bibitem{Ser10} I. Serban, B. B\'{e}ri, A. R. Akhmerov, and C. W. J. Beenakker, Phys. Rev. Lett. \textbf{104}, 147001 (2010).
\bibitem{Chu11} S. B. Chung, X.-L. Qi, J. Maciejko, and S.-C. Zhang, Phys. Rev. B \textbf{83}, 100512(R) (2011).
\bibitem{Liu11} C.-X. Liu and B. Trauzettel, Phys. Rev. B \textbf{83}, 220510(R) (2011).
\bibitem{Str11} G. Str\"{u}bi, W. Belzig, M.-S. Choi, and C. Bruder, Phys. Rev. Lett. \textbf{107}, 136403 (2011).
\bibitem{Li12} J. Li, G. Fleury, and M. B\"{u}ttiker, Phys. Rev. B \textbf{85}, 125440 (2012).
\bibitem{Ber12} B. B\'{e}ri, Phys. Rev. B \textbf{85}, 140501(R) (2012).
\bibitem{Teo10}  J. C. Y. Teo and C. L. Kane, Phys. Rev. B \textbf{82}, 115120 (2010).
\bibitem{Wie13} B. J. Wieder, F. Zhang, and C. L. Kane, arXiv:1302.2113.
\bibitem{Akh11} A. R. Akhmerov, J. P. Dahlhaus, F. Hassler, M. Wimmer, and C. W. J. Beenakker, Phys. Rev. Lett. \textbf{106}, 057001 (2011).
\bibitem{Alt97} A. Altland and M. R. Zirnbauer, Phys. Rev. B \textbf{55}, 1142 (1997).
\bibitem{Bee11} C. W. J. Beenakker, J. P. Dahlhaus, M. Wimmer, and A. R. Akhmerov, Phys. Rev. B \textbf{83}, 085413 (2011).
\bibitem{Ful11} I. C. Fulga, F. Hassler, A. R. Akhmerov, and C. W. J. Beenakker, Phys. Rev. B \textbf{83}, 155429 (2011).
\bibitem{note1} The symmetry requirement \eqref{CDsymmetry} and the requirement of gauge invariance \eqref{gaugeinvariance} constrain the functional form of the pair potential in the Bogoliubov-De Gennes Hamiltonian, but some freedom is left. An alternative gauge invariant \textit{d}-wave pair potential was introduced by O. Vafek, A. Melikyan, M. Franz, and Z. Te\v{s}anovi\'{c}, Phys. Rev. B \textbf{63}, 134509 (2001).
\bibitem{parameters} Model parameters for the superconductor are $\mu=t$, $\Delta_{0}=0.4\,t$ for the chiral \textit{p}-wave pairing, $\mu=t$, $\Delta_{0}=0.6\,t$, $\alpha_{\rm so}=0.2\,t/p_{\rm F}$ for the helical \textit{p}-wave pairing, and $\mu=t$, $\Delta_{0}=0.4\,t$ for the chiral \textit{d}-wave pairing.
\bibitem{Wim09} M. Wimmer and K. Richter, J. Comp. Phys. \textbf{228}, 8548 (2009).
\bibitem{Sac11} B. Sac\'{e}p\'{e},	 J. B. Oostinga, J. Li, A. Ubaldini, N. J. G. Couto, E. Giannini, and A. F. Morpurgo, Nature Comm. \textbf{2}, 575 (2011).
\bibitem{Wil12} J. R. Williams, A. J. Bestwick, P. Gallagher, S. S. Hong, Y. Cui, A. S. Bleich, J. G. Analytis, I. R. Fisher, and D. Goldhaber-Gordon, Phys. Rev. Lett. \textbf{109}, 056803 (2012).
\bibitem{Vel12} M. Veldhorst, M. Snelder, M. Hoek, T. Gang, X. L. Wang, V. K. Guduru, U. Zeitler, W. G. van der Wiel, A. A. Golubov, H. Hilgenkamp, and A. Brinkman, Nature Mat. \textbf{11}, 417 (2012).
\bibitem{Sau10} J. D. Sau, R. M. Lutchyn, S. Tewari, and S. Das Sarma, Phys. Rev. Lett. \textbf{104}, 040502 (2010).
\bibitem{Ali10} J. Alicea, Phys. Rev. B \textbf{81}, 125318 (2010).
\bibitem{Dah12} J. P. Dahlhaus, M. Gibertini, and C. W. J. Beenakker, Phys. Rev. B \textbf{86}, 174520 (2012).
\bibitem{Pei33} R. E. Peierls, Z. Phys. \textbf{80}, 763 (1933).
\end{thebibliography}
\end{document}